\documentclass[two column,prb]{revtex4}
\usepackage{bm}
\usepackage{graphics}
\usepackage{graphicx}
\usepackage{amsfonts}
\usepackage{amsmath}
\usepackage{amssymb}
\usepackage{graphicx}
\usepackage{color}

\begin{document}

\preprint{HEP/123-qed}
\title[Short title for running header]{Scanning probe microscopy and field emission schemes for studying electron emission from polycrystalline diamond}
\author{Oksana Chubenko$^{1,2}$}
\author{Stanislav S. Baturin$^2$}
\email{s.s.baturin@gmail.com}
\author{Sergey V. Baryshev$^2$}
\email{sergey.v.baryshev@gmail.com} \affiliation{$^1$Department of
Physics, The George Washington University, 725 21st St. NW,
Washington, DC 20052, USA
\\
$^2$Euclid TechLabs, 365 Remington Blvd., Bolingbrook, IL 60440,
USA}

\begin{abstract}
The letter introduces a diagram that rationalizes tunneling atomic
force microscopy (TUNA) observations of electron emission from
polycrystalline diamonds as described in recent publications
[1,2]. The direct observations of electron emission from grain
boundary sites by TUNA could indeed be evidence of electrons
originating from grain boundaries under external electric fields.
At the same time, from the diagram it follows that TUNA and field
emission schemes are complimentary rather than equivalent for
results interpretation. It is further proposed that TUNA could
provide better insights into emission mechanisms by measuring the
detailed structure of the potential barrier on the surface of
polycrystalline diamonds.
\end{abstract}

\maketitle

The question is yet to be answered of why synthetic
polycrystalline diamonds (micro-, nano-, or ultra-nano-crystalline
diamond, abbreviated MCD, NCD, UNCD) containing a large amount of
the carbon $sp^2$ phase are such excellent electron field
emitters. These diamonds have a low threshold (turn-on) electric
field $\sim$10 MV/m and yield significant current densities. A
powerful approach, called the graphitic patch model, to explain
this behavior was attempted by Cui, Ristein and Ley [3]. It first
originated to plausibly explain sub-bandgap photoelectric emission
in single-crystal diamond. The main idea was that the surface
always has small carbonic (graphitic) phase patches (electron
emitters). The work function (4.6 eV) is reduced when the property
of negative electron affinity (NEA, which can be as low as $-$1.3
eV) is induced on the surrounding diamond host surface.
Experimentally, the potential barrier [4] of a patch can be as low
as 3.0 eV. The model showed excellent agreement with experiments.
While the existence of carbon patches on the surface of a
single-crystal diamond could be questioned (signatures seen from
indirect spectroscopic measurements), in polycrystalline diamonds
the carbon $sp^2$ phase is a separation interlayer between diamond
micro- or nano-crystallites (or grains), and can be directly
imaged by transmission electron microscopy in large amounts. The
carbon interlayer is also called the grain boundary (GB). Recent
observations [1,2] have demonstrated that tunneling electron
emission originates from GBs. The observations were made by a
specialty atomic force microscope equipped with tunneling current
measurement capability, abbreviated as TUNA. The authors
hypothesized that TUNA measurements should be representative in
the conventional field emission case, meaning that GBs emitting in
the TUNA scheme should be emitting sites in a field emitter based
on a polycrystalline diamond. Nevertheless, they did not provide a
straightforward explanation how exactly TUNA represents the field
emission mechanism. In this letter, we present a diagram bridging
the TUNA scheme results and the conventional field emission
scheme. The diagram is based on the graphitic patch model and the
usual electrostatics.

Panel (a) in the figure illustrates a case in which a GB of a
lateral size of $\lesssim$1 nm is brought in contact with an
intrinsic diamond grain of lateral size $\geq$10 nm. The Fermi
level ($\varepsilon_f$) in diamond is higher on the energy scale
compared to that in carbon, which should result in a contact
potential difference and upward band bending. As explained in
great detail in Ref.[3], upon hydrogen termination of diamond,
carbon supplies electrons which are liberated close to the triple
graphite/diamond/vacuum interface having a reduced potential
barrier suppressed by the induced NEA. When the graphitic patch
size tends to an extremely small value at the limit, much smaller
than the surrounding diamond host, the whole patch should reduce
the potential barrier by a number close to the NEA value of
diamond (from 4.6 eV to 3.0 eV as found in Ref.[3]). The barrier
height/emission threshold of 3.0 eV was later confirmed in both
photo- and field-emission experiments conducted for the same
undoped polycrystalline diamond [5]. In $sp^2$ carbon, electrons
fill the full band below its Fermi level and are ready for
resonant tunneling.

From the description of the TUNA tool, it follows that it is
essentially a Kelvin probe in which the distance between the
sample and the reference counter electrode ($Pt-Ir$) can be
controlled with superfine precision and accuracy, and the surface
topography can be recorded along with tunnel current maps. The
panel (b) in the figure illustrates the energy balance diagram in
the Kelvin probe method. For two separated conductive materials,
vacuum levels are equal while Fermi levels are located below the
vacuum level by values equal to respective work functions. When
two separated samples are connected at the back through a voltage
source, Fermi levels can be moved up and down with respect to each
other. Electric potential difference is induced when energy
position of the Fermi levels differ. $\Delta_{EP}$ is maximal and
equal to 2.5 eV=5.5 eV$-$3.0 eV when the Fermi levels in carbon
and $Pt-Ir$ are equilibrated, while it is zero when vacuum levels
are equilibrated (i.e. when the electrodes are disconnected);
$\Delta_{EP}$ varies between 0 eV and 2.5 eV with applied voltage
$V$. In the TUNA experiment, voltage spanning +1 mV to +1 V was
applied to the reference $Pt-Ir$ electrode, meaning its Fermi
level was always lower than that of carbon by 1 meV to 1 eV. Thus,
electrons flow from the sample to the $Pt-Ir$ electrode, i.e. in
the direction coincidentally equivalent to the field emission
case. Increase in the reading current should be observed when
increasing the voltage applied to the counter electrode. Such
increase was indeed observed. Current grew by a factor of
$\sim$100 (few pA to few 100's of pA) when the voltage was varied
by a factor of $\sim$1,000 (1 mV to 1 V). This dependence is
entirely different compared to the field emission mechanism of
tunneling through a triangle barrier described by the
Fowler-Nordheim formula. The barrier in the TUNA scheme has a
different configuration. At a short range of a few nm, an electron
experiences the image potential ($IP$), which grows from the Fermi
level to the global asymptotic vacuum level as
\[
IP(x)=-e\cdot\frac{e}{16\cdot \pi \cdot \varepsilon_0 \cdot x}
\text{, (1)}
\]
where $e$ is the electron charge and $\varepsilon_0$ is the vacuum
permittivity. In this case, we set the image potential growth
scale to 4 nm. At distances farther than 4 nm, the image potential
level is less than 0.1 eV below the vacuum level, meaning these
potentials merge and become indistinguishable. With the tip placed
at $d$=1 nm, the resulting barrier will be formed and modified
with the sum of the $IP$'s of the two electrodes (intercepted at a
scale where the $IP$ is lower than the asymptotic vacuum level)
and the $EP$ (depends on the original Fermi level positions in the
electrodes prior to connecting, and the voltage applied between
the electrodes). An exemplary resulting barrier shown in the panel
(c) is for the case with +1 V applied to the $Pt-Ir$ electrode:
$\Delta_{EP}$=1.5 eV and $\varepsilon_f$ of $Pt-Ir$ is 1 eV below
that of the carbon patch. Here, the zero energy level is set as
the vacuum level on the carbon contact side. The resulting barrier
is the sum of $IP_1=IP^{carbon}$,
$IP_2=IP^{Pt-Ir}+(\varepsilon_f^{Pt-Ir}-\varepsilon_f^{carbon}-V)$,
and electric potential
$\Delta_{EP}=(\varepsilon_f^{Pt-Ir}-\varepsilon_f^{carbon}-V)\cdot
\frac{x}{d}-(\varepsilon_f^{Pt-Ir}-\varepsilon_f^{carbon}-V)$. In
this representation current growth, measured by TUNA, is simply
proportional to the number of states in $Pt-Ir$ available for
tunneling, as the Fermi level in the counter electrode slides down
with increasing applied voltage. There is some likelihood that the
$I$ versus $V$ behavior is non-monotonic because simultaneously
more states in $Pt-Ir$ open up and the resulting barrier height
and width are reduced. Carrier depletion in the carbon patch/GB
could be another effect.

The TUNA results placed into context of the graphitic patch model
make sense. In turn, observation of electron emission in
insulating polycrystalline diamond can be only explained by the
presence of conductive inclusions, i.e. GBs, because electrons in
diamond grains are bound. If TUNA experiments were carried out
away from the sample edges, it suggests GBs are electrically
connected via some mechanism. The observed steady increase in
surface emitting area from undoped MCD, to NCD, to UNCD (Fig.1 in
Ref.[1]) should be caused by the increased ratio of graphitic
$sp^2$ phase (GB) to diamond $sp^3$ phase (grain). In terms of
emitting area, $N$-doped NCD did not differ from UNCD since
nitrogen is incorporated into GBs changing their size
insignificantly compared to the size of surrounding grains.
Further, the surge in the emitting area in $P$- and $B$-doped MCD
films can be explained by actual grain doping that takes place so
that GBs and grains become conductive and have impurity state
bands able to supply electrons.

Similar situation seems to take place in case of scanning
tunneling microscopy (STM) measuring electron emission properties
of UNCD with embedded metallic nanoparticles [6]. From actual
field emission data, it is seen that the ratio of the barrier
height to the field enhancement factor, $\varphi^{3/2}/\beta$, in
the exponent of the Fowler-Nordheim equation almost did not
change, while the current density scaled-up significantly. It
means that it is quite possible that enhanced emission properties
are due to increased electron density introduced by the metallic
nanoparticles (especially in the case of Au) and a larger $sp^2$
phase distribution network boosted by the Au implantation.
Additional confirmation by STM that Au nanoparticles produce
measurable current alongside with GBs suggests that Au particles
have a vast amount of electrons that can readily tunnel under a
barrier produced by the sample surface and the counter electrode
tip of an STM tool, situation equivalent to TUNA.

\begin{figure*}[t]
\includegraphics[height=4.7cm]{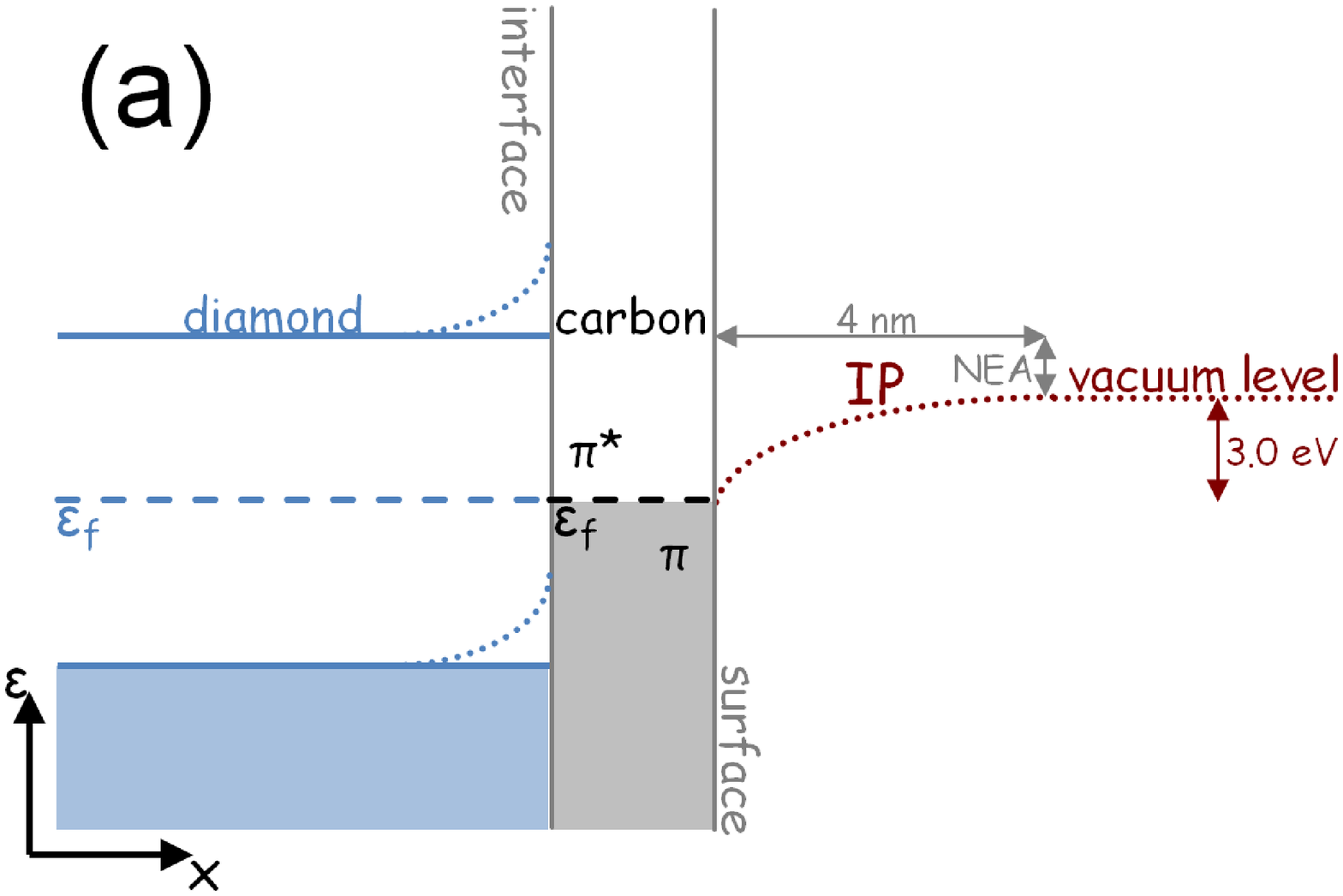}
\hspace{0.3cm}
\includegraphics[height=4.7cm]{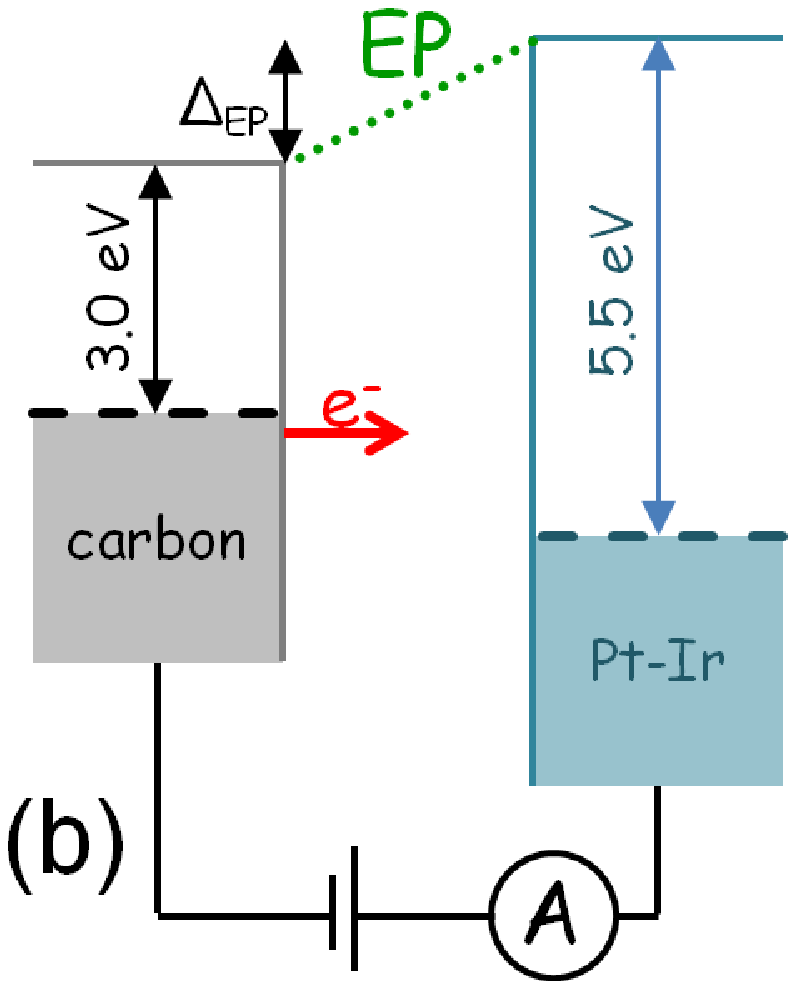}
\hspace{0.3cm}
\includegraphics[height=4.7cm]{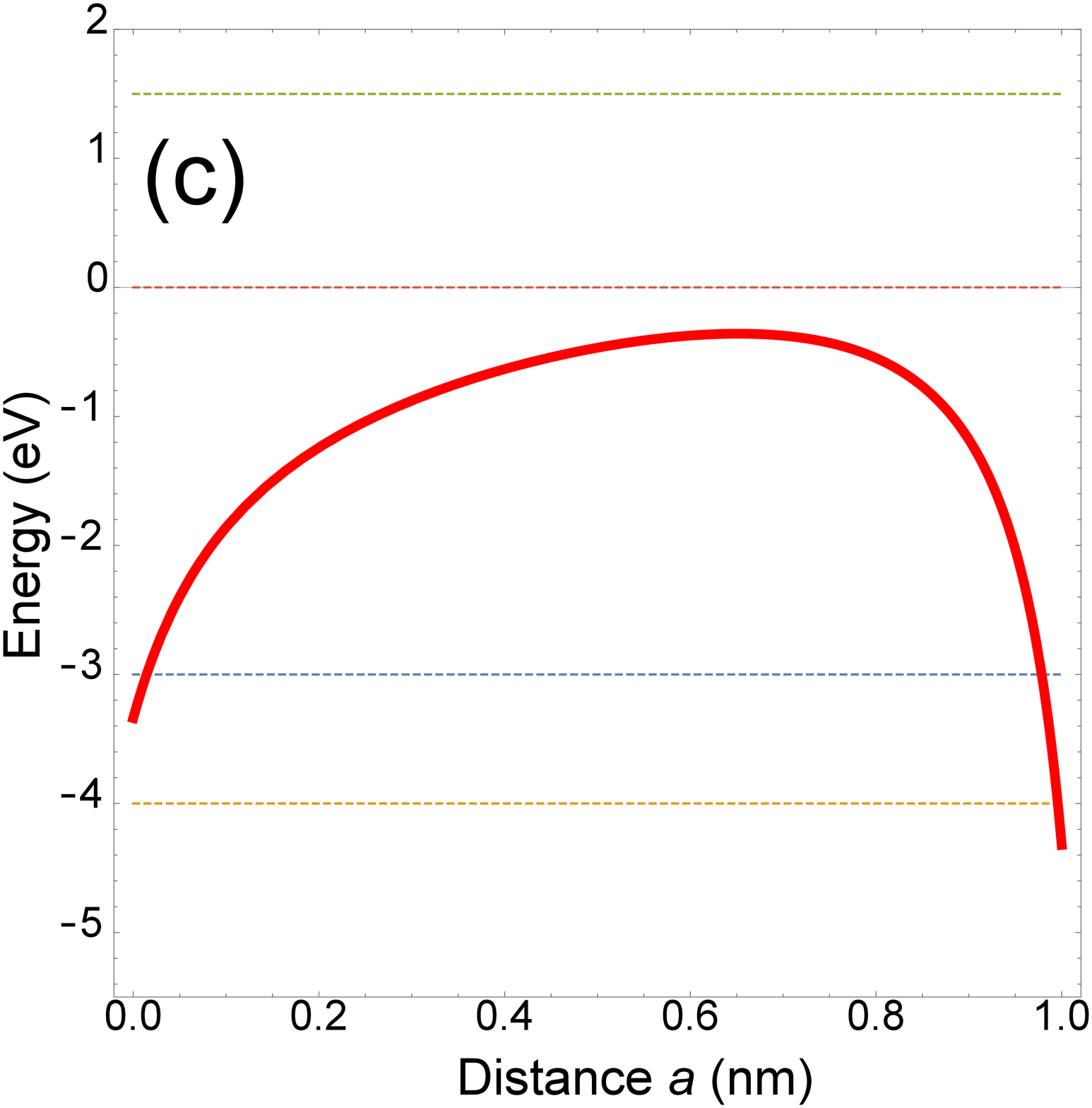}
\caption{\textbf{Basic diagram showing (a) an $sp^2$ GB adjacent
to an intrinsic diamond crystallite/grain; (b) measurement diagram
of the TUNA setup; (c) an example of a resulting potential barrier
formed between the carbon patch (i.e. GB) connected to ground and
the TUNA $Pt-Ir$ reference tip biased at +1 V. The zero energy
level represents the vacuum level on the carbon contact side.}}
\end{figure*}

Another important consequence of the TUNA observations is that it
rules out the assumption that the field enhancement factor $\beta$
of pure geometrical nature plays significant role in field
emission, because electrons originate from GB topographical
valleys. One way to design an alternative $\beta$ was proposed by
Robertson [7], who illustrated a mechanism of self-focusing
electric field lines at the hydrogen-terminated/unterminated
discontinuities on the ideal planar surface giving rise to
electric gradients $\sim$1 GV/m. Yet another possible way to
introduce a $\beta$ of quasi-geometrical nature is as follows. In
any undoped polycrystalline diamond or in incorporated NCD and
UNCD, grains stay insulating. This means that nothing prevents
electric field from penetrating through dielectric grains and
focusing on conductive GBs (roughly of a shape of a blade as can
be assumed from top-view and cross-section TEM images [8,9]). Such
effect would be $\varepsilon_r=5.7$ (relative permittivity of
diamond) times smaller as compared to the case of the identical
conductive blades placed in an external electric field in vacuum
with $\varepsilon_r=1$. Overall such effect is modest, and field
enhancement should not go over $10-20$. Nevertheless, it is enough
to reasonably account for $\beta>1$ found in experiments - see for
instance updoped unterminated case for MCD in Ref.[5] showing
$\beta=18$. In polycrystalline diamonds where actual doping of
grains can be achieved, such effect would vanish because the whole
surface becomes conductive and the electric lines would terminate
on the surface perpendicularly, meaning $\beta\rightarrow 1$.

This means the potential barrier has to be the key to field
emission in this class of materials. Improved emission properties
of polycrystalline diamonds, especially upon doping/incorporation
and termination/functionalization, mean the barrier height of the
GB may reduce further from 3.0 eV to a lower number ($\lesssim$1
eV). Mechanisms responsible could be the significant stress
present in polycrystalline films [10], changing the absolute
energy position of the Fermi level, and/or structural or chemical
reconfigurations. As an example of the latter, in amorphous
hydrogenated carbon nitride, a sister system to the GB material
[11], barriers as low as 0.7 eV were observed [12]. The described
and/or some other mechanisms acting synergistically could produce
a surface barrier of a modified shape that enhances the tunneling
current significantly.

To conclude, while scanning probe microscopy and field emission
schemes are not directly equivalent, tunneling measurements help
better understand and interpret field emission data taken for a
parent (undoped, unterminated) polycrystalline diamond and field
emission property changes upon bulk and surface modification of
the same parent material. In various CVD diamond types with varied
doping level/type and grain/GB size, taking advantage of the
unique TUNA capabilities one could measure the surface barrier
structure and Fermi level position using different potentials and
polarities of the reference tip, inter-electrode separation and
additional sample excitations. The electron supply function of a
GB, i.e. electrons per emitting site and emission rate, density of
states (cm$^{-3}\times$eV$^{-1}$), and conductivity mechanism
through GBs could be also determined.

We thank Paul Schoessow for his help with the manuscript.


\begin{thebibliography}{}
\bibitem{4} To avoid any confusion, in what follows we use a more general
term potential barrier to describe the penalty for an electron to
escape into vacuum.
\end{thebibliography}
\end{document}